\documentstyle[preprint,eqsecnum,aps,epsfig]{revtex}
\tightenlines

\begin{document}
\draft
\preprint{HUPD-0111, KUNS-1726}

\title{Chiral phase transition of bulk Abelian gauge theories in the Randall-Sundrum brane world}

\author{Hiroyuki Abe$^1$, Kenji Fukazawa$^2$ and Tomohiro Inagaki$^3$\\\ }

\address{$^1$ Department of Physics, Kyoto University,
Kyoto, 606-8502, Japan; \\
Department of Physics, Hiroshima University,
Higashi-Hiroshima, 739-8526, Japan\\
E-mail: abe@gauge.scphys.kyoto-u.ac.jp\\ }
\address{$^2$ Kure National College of Technology,
Kure, 737-8506, Japan\\
E-mail: fukazawa@kure-nct.ac.jp\\}
\address{$^3$ Information Media Center, 
Hiroshima University,
Higashi-Hiroshima, 739-8521, Japan\\
E-mail: inagaki@hiroshima-u.ac.jp\\\ }

\maketitle

\begin{abstract}
The chiral phase transition of strong-coupling Abelian gauge theories 
is investigated in the brane world.
It is assumed that gauge boson propagates in an extra dimension, i.e. 
bulk gauge theories.
The phase structure is analytically evaluated by using the low-energy 
effective theories. We also numerically solve the ladder Schwinger-Dyson 
equation for the full fermion propagator including Kaluza-Klein (KK) 
excitation modes of the gauge field. It is found that 
the chiral phase transition is of the second order, and the critical 
value of the coupling constant is obtained. The extra dimension has a 
large influence on the chiral phase transition for the Randall-Sundrum 
(RS) brane world. It is studied how the number of KK modes affect the 
chiral phase transition.
\end{abstract}

\narrowtext

\section{Introduction}

The idea of compact extra dimensions has been proposed to solve the 
hierarchy problem between the Planck scale and the electroweak scale
 in Ref.\cite{Ant,Ant2,Ark1}. To avoid fine tuning the radius
of compact extra dimensions, Randall and Sundrum introduced Minkowski
branes in five-dimensional anti-deSitter space-time which is a static
solution to Einstein's equation \cite{Ran}. At the beginning, it has
usually been assumed that the standard model particles are confined in the 
Minkowski brane while the graviton can propagate in extra dimensions.
In the present paper we study the influence of bulk gauge field on
chiral phase transition. 

A possible mechanism to confine the standard model particles in the
brane has been found by constructing a bulk standard model \cite{Cha}. In
the bulk standard model KK excitation modes \cite{KK} of the
standard model particles appear in the brane. A variety of low energy
phenomena have been studied in the bulk standard model. One of the most
interesting phenomena is found in spontaneous electroweak symmetry 
breaking of gauge theories. It may play a decisive role in
the standard model and the grand unified theories. The dynamical mechanism 
of electroweak symmetry breaking has been reexamined within 
brane world models \cite{Dob,Chen,Kob,Abe1,Riu,Abe2}. 
It has been pointed out that the
exchange of KK modes enhances dynamical symmetry breaking. A
possibility of the top quark condensation is revived along 
with the brane world scenario \cite{ArkPRD,Has}. Dynamical symmetry
breaking has been evaluated in higher-dimensional compact space-time,
and then the result is reduced to the four-dimensional branes. But 
the role of the brane particle in symmetry breaking 
is not clear in previous studies. The brane particle may have a
non-negligible effect on symmetry breaking \cite{Gor}.

There are a variety of models for the brane world. In many models it is
assumed that compact extra dimensions exist. It has been shown that
the finite size effect and the boundary condition can dramatically
change the phase structure of dynamical symmetry
breaking \cite{Ina2,Fuk}. The RS brane world has two Minkowski branes in 
five-dimensional negative curvature space-time, $AdS^5$. The space-time 
curvature also has an important effect on the phase structure 
of dynamical symmetry breaking \cite{Ina1,Ina4}. Especially in negative 
curvature
space-time it has been shown that chiral symmetry is always broken
down for models with four-fermion interactions if the space-time
dimensions are less than four \cite{Ina2,Ina1,Ina3}. Therefore, we
launched a plan to study dynamical symmetry breaking in the RS brane
world using the induced four-dimensional Lagrangian 
from five-dimensional gauge theory.
In Refs. \cite{Abe1} and \cite{Abe2} the structure of chiral
symmetry breaking has been studied in the bulk fermion models. It is pointed
out that the dynamical fermion mass is much smaller than the Planck scale.

In this paper we consider five-dimensional Abelian gauge
theories as a simple model where chiral symmetry is broken down
dynamically. The structure of chiral phase transition is investigated in 
the bulk gauge theories. In \S 2 we assume
that gauge boson propagates in an extra dimension, i.e. 
bulk gauge theories. The explicit expressions of the induced 
four-dimensional Lagrangian are obtained.
In \S 3 we construct low-energy effective theories with local four-fermion
interactions. Using the bifurcation method phase structure is
analytically investigated in the effective theories. In \S4
we numerically solve the ladder Schwinger-Dyson equation including KK modes 
of the gauge field.

\section{Induced four-dimensional Lagrangian of bulk gauge theories}

Throughout this paper we assume an existence of five-dimensional bulk gauge
fields, while the fermion fields are confined in the brane.
We consider the Abelian gauge theories with massless fermions. Since we are interested in KK excitations for gauge fields,
the KK excitations for graviton are assumed to have no serious effect
on chiral symmetry breaking, and thus are ignored.
We take account of KK excitations for gauge fields only, and construct the induced four-dimensional Lagrangian from the five-dimensional theory in the brane. 

If the gauge fields are confined in the brane, an extra dimension does not contribute to gauge interactions. The Lagrangian for Abelian gauge theories has a familiar form:
\begin{equation}
{\cal L\/}^{(4)}
= \frac{1}{2} A_\mu \left[ 
    \eta^{\mu \nu} \partial^2 - \left( 1-\xi \right) \partial^\mu \partial^\nu
    \right] A_\nu
+ \bar{\psi} \left( i \partial_\mu + e A_\mu \right) \gamma^\mu \psi.
\label{lag:org}
\end{equation}
It has been known for a long time that a fermion acquires the
dynamical mass  and that chiral symmetry is broken down for strong
coupling, $\alpha=e^2/(4\pi) > \pi/3$ \cite{BJW,MN}. 

Below, we consider the bulk gauge field which propagates in an extra
dimension. We then construct the induced four-dimensional Lagrangian.

\subsection{Bulk Abelian gauge theories in flat extra dimension}

We first study a flat space-time with a single orbifold extra dimension, $M^4\otimes S^1/Z_2$. The extra dimension is assumed to be compactified as a circle with radius $R$. According to the $S^1/Z_2$ invariance, we adopt the following boundary conditions,
\begin{eqnarray}
A_{\mu}(x,y+1)=A_{\mu}(x,y), A_{\mu}(x,y)=A_{\mu}(x,-y),
\\
A_{4}(x,y+1)=A_{4}(x,y), A_{4}(x,y)=-A_{4}(x,-y),
\end{eqnarray}
where $x$ is a coordinate on four-dimensional Minkowski surfaces, $y$ is
a coordinate in an extra dimension normalized by $2\pi R$ and
$\mu=0,1,2,3$. In the branes, $y=y^\ast$ $(=0)$, so these conditions simplify
\begin{eqnarray}
\partial_{4}A_{\mu}(x,y=y^\ast)=0,
\\
A_{4}(x,y=y^\ast)=0.
\end{eqnarray}
The $Z_2$ symmetry prevents the gauge field $A_{4}$ from having
non-vanishing value in the brane. The field $A_{4}$ cannot couple 
with the brane particle directly. And furthermore, we apply the gauge
fixing conditions where the extra component of the gauge fields,
$A_{4}$, is decoupled from $A_{\mu}$. (See Appendix A.) Thus, the fifth 
component of the gauge fields disappears from the Lagrangian 
on the brane \cite{Kug,Han}.

By adopting the periodic boundary condition and the $Z_2$ invariance,
 the bulk gauge field is decomposed into Fourier modes:
\begin{eqnarray}
A_\mu (x,y) = \frac{1}{\sqrt{2\pi R}}
  \sum_n A_\mu^{(n)}(x)\mbox{cos}(2\pi ny) .
\label{modeex:torus}
\end{eqnarray}

The induced four-dimensional Lagrangian is described by 
\begin{eqnarray}
{\cal L\/}^{(4)}
&=&\int^{1/2}_0 dy\, 
    \left[ -\frac{1}{4} F^{MN}F_{MN} + {\cal{L}}_{\rm{G.F.}} +{\cal{L}}_{\rm {F}}\delta(y-y^\ast) \right]
    \nonumber \\
&=& \frac{1}{2} \sum_{n} A_\mu^{(n)} \left\{
  \eta^{\mu \nu} \left[ \partial^2 + \left(\frac{n}{R}\right)^2 \right]
  - \left( 1-\xi \right) \partial^\mu \partial^\nu
  \right\} A_\nu^{(n)}
  \nonumber \\ 
&&  + \bar{\psi} \left( i \partial_\mu + e\sum_{n}A_\mu^{(n)} \right)
  \gamma^\mu \psi ,
\label{lag:torus}
\end{eqnarray}
where ${\cal{L}}_{\rm{G.F.}}$ represents the gauge fixing term and 
${\cal{L}}_{\rm {F}}$ is the Lagrangian for massless fermion.
The contribution from the extra dimension is imposed in the KK modes, $A_\mu^{(n)}(x)$, with mass $n/R$.

\subsection{Bulk Abelian gauge theories in warped extra dimension}

The existence of the brane naturally leads to the curved extra
dimensions. L. Randall and R. Sundrum found the curved space-time with
Minkowski branes as a static solution of the Einstein equation \cite{Ran}. The model has {\it two} Minkowski branes at the $S^1/Z_2$ orbifold fixed points,
$y^\ast=0, 1/2$, and the {\it AdS$_5$} between these branes:
\begin{eqnarray}
ds^2 \equiv G_{MN} dx^M dx^N 
     = e^{-2kb_0|y|}\eta_{\mu\nu}dx^\mu dx^\nu - b_0^2dy^2, 
\label{metric}
\end{eqnarray}
where $k \sim M_{pl}$ is the gravity scale, and $b_0^{-1} \sim k/24\pi$
is the RS compactification scale. The exponential factor, $e^{-2kb_0|y|}$,
in Eq.(\ref{metric}) is called the warp factor. One of the most important results of the RS model is the warp factor, $e^{-\frac{1}{2}kb_0}$, suppression of the KK mode's mass for the bulk fields.

Following the procedure developed in Ref. \cite{Cha,Gol}, we derive the mode expansion of the bulk gauge fields in RS space-time such that 
\begin{eqnarray}
A_\mu (x,y) = 
  \sum_n A_\mu^{(n)}(x){\hat{\chi}_n (y)},
\label{modeex:rs}
\end{eqnarray}
and the mode functions are
\begin{eqnarray}
{\hat{\chi}_n (y)}
  = \frac{\sqrt{kb_0} e^{kb_0|y|}}{N_n}
    \left[
      J_1 \left( \frac{M_n}{k} e^{kb_0|y|} \right)
    - \frac{J_0 \left( M_n/k \right)}{Y_0 \left( M_n/k \right)} 
      Y_1 \left( \frac{M_n}{k} e^{kb_0|y|} \right)
    \right] .
\label{eq:chi}
\end{eqnarray}
where the KK mode's mass $M_n$ is given by the solution of 
\begin{eqnarray}
 J_0 \left( \frac{M_n}{k} \right) 
 Y_0 \left( \frac{M_n}{k} e^{\frac{1}{2} kb_0} \right) =
 Y_0 \left( \frac{M_n}{k} \right)
 J_0 \left( \frac{M_n}{k} e^{\frac{1}{2} kb_0} \right),
\label{eq:KK}
\end{eqnarray}
and $N_n$ is a normalization factor:
\begin{eqnarray}
 N_n^2=\left.z^2\left[J_1 \left( \frac{M_n}{k}z \right)
       -\frac{J_0 \left(M_n/k \right)}
             {Y_0 \left( M_n/k \right)} 
       Y_1 \left( \frac{M_n}{k}z \right) \right]^2
       \right|^{Z=\exp(kb_0/2)}_{Z=1} \ \ .
\end{eqnarray}

For $kb_0 \gg 1$ and $M_n \ll k$ the KK mode's mass behaves
asymptotically like \cite{Pom}
\begin{equation}
  M_n \sim \left(n-\frac{1}{4}\right)\pi k e^{-\frac{1}{2} kb_0}.
\end{equation}
In this case, $n$ dependence of the mode function $\hat{\chi}_{n}(y)$
is negligible in the brane $y=1/2$. It behaves as \cite{Kub}
\begin{equation}
  \hat{\chi}_{n}(1/2) \sim \sqrt{kb_0}.
\label{chi:app}
\end{equation}

Performing the KK modes decomposition (\ref{modeex:rs}), the induced 
four-dimensional Lagrangian for a free gauge field reads
\begin{eqnarray}
{\cal L\/}^{(4)}_{\rm gauge} 
&=& \int_0^{\frac{1}{2}} dy\, 
    \left[ -\frac{1}{4} \sqrt{-G} F^{MN}F_{MN} + {\cal{L}}_{\rm{G.F.}} \right]
    \nonumber \\
&=& \frac{1}{2} \sum_{n} A_\mu^{(n)} \left[ 
    \eta^{\mu \nu} \left( \partial^2 + M_n^2 \right)
    - \left( 1-\xi \right) \partial^\mu \partial^\nu
    \right] A_\nu^{(n)}.
\end{eqnarray}
As is mentioned in the previous subsection, the fifth component of the gauge fields disappears by the $Z_2$ symmetry and the gauge fixing condition.

Thus, we find the four-dimensional Lagrangian with massless fermion,
\begin{eqnarray}
{\cal L\/}^{(4)}
 = \frac{1}{2} \sum_{n} A_\mu^{(n)} \left[
 \eta^{\mu \nu} \left( \partial^2 + M_n^2 \right)
 - \left( 1-\xi \right) \partial^\mu \partial^\nu
 \right] A_\nu^{(n)}
\nonumber \\ 
 + \bar{\psi} \left( i \partial_\mu + eA_\mu^{(0)} \right) \gamma^\mu \psi
 + e \sum_{n\ne 0} \hat{\chi}_n (y^\ast) \bar{\psi} A_\mu^{(n)} 
 \gamma^\mu \psi .
\label{lag:rs}
\end{eqnarray}
In the warped extra dimension the contribution from the extra dimension
appears in the coupling constant of the interactions between the KK
modes and the fermion. From Eq. (\ref{chi:app}) the KK modes
have strong coupling in the brane for large $kb_0 (\gg 1)$ \cite{Cha}.

In the following section we study the structure of chiral symmetry breaking starting from the Lagrangian (\ref{lag:org}), (\ref{lag:torus}) and (\ref{lag:rs}).

\section{Local four-fermi approximation}\label{sec3}

As a first step to investigate the phase structure, we assume that the
KK mode's mass $M_n$ is on the order of the cut-off scale $\Lambda$
analytically to deal with chiral symmetry breaking. If all the KK excited modes are heavier enough, terms with KK modes in the Lagrangian (\ref{lag:torus}) and (\ref{lag:rs}) reduce to four-fermion interactions. After integrating out the KK excited modes, we obtain
\begin{equation}
{\cal L}_{\rm eff} 
    = -\frac{1}{4} F^{(0)\mu \nu}F^{(0)}_{\mu \nu} 
      +\bar{\psi} \left( i\partial_\mu +eA_\mu^{(0)} \right) \gamma^\mu \psi
      -\frac{1}{2} 
       G \left( \bar{\psi} \gamma^\mu \psi \right) 
         \left( \bar{\psi} \gamma_\mu \psi \right) .
\label{lag:ff}
\end{equation}
The four-fermion coupling constant $G$ is given by
\begin{eqnarray}
G = &\displaystyle \sum_{n \ne 0}^{N_{KK}}\frac{4\pi\alpha R^2}{n^2} 
\ \stackrel{N_{KK}\rightarrow \infty}{\longrightarrow}\  \frac{2 \pi^2 \alpha R^2}{3}\ , 
& \mbox{     (for flat extra dimension)}
\label{G1:GNJL}
\\
G = &\displaystyle \sum_{n \ne 0}^{N_{KK}} \frac{4\pi\alpha}{M_n^2} \hat{\chi}_n (y^\ast)^2\ .
& \mbox{     (for warped extra dimension)}
\label{G2:GNJL}
\end{eqnarray}
If the gauge field is confined in the brane, the four-fermion interaction
term is not generated, i.e. $G=0$. 

Using the Fierz transformation, the four-fermion interaction terms in (\ref{lag:ff}) are rewritten as
\begin{eqnarray}
\cal{L}_{\rm eff} 
  &=& -\frac{1}{4} F^{(0)\mu \nu}F^{(0)}_{\mu \nu} 
      +\bar{\psi} \left( i\partial_\mu +eA_\mu^{(0)} \right) \gamma^\mu \psi
      +\frac{1}{2} 
       G \left[ \left( \bar{\psi} \psi \right)^2 + 
                \left( \bar{\psi} i\gamma_5 \psi \right)^2 \right] ,
\label{GNJLLag}
\end{eqnarray}
where we ignore the tensor type interactions. Eq. (\ref{GNJLLag}) is nothing but the  gauged Nambu-Jona-Lasinio (NJL) model. A variety of methods have been developed to evaluate the phase structure in the gauged NJL model, (see \cite{Yam,Mir}).
According to the bifurcation method, the critical line in the $\alpha$-$G$ plane is analytically obtained \cite{App,Kon}
\begin{equation}
\frac{G_c \Lambda^2}{4\pi^2}=\frac{1}{4}\left(1+\sqrt{1-\frac{3\alpha_c}{\pi}}\right)^2\ ,\ \ \  \left(\alpha < \frac{\pi}{3}\right)
\label{cl:GNJL}
\end{equation}
where $G_c$ and $\alpha_c$ denote the critical values of $G$ and $\alpha$ respectively. For $\alpha < \pi /3$ and $G > G_c$, the fermion mass 
function $B(p^2)$ behaves as
\begin{equation}
  B(p^2)=\frac{\sqrt{1-\omega^2}}{\omega}\frac{\mu^2}{\sqrt{p^2}} 
  \mbox{sinh}\left(
    \frac{\omega}{2}\mbox{ln}\frac{p^2}{\mu^2}+\mbox{tanh}^{-1}\omega
  \right) ,
\label{mass:GNJL}
\end{equation}
where $\omega=\sqrt{1-3\alpha/\pi}$ and $\mu$ is given by
\begin{equation}
  \mu=\Lambda \exp \left[
    -\frac{1}{\omega}\mbox{tanh}^{-1}\left(
      \frac{\omega/2}{-1/4-\omega^2/4+G \Lambda^2/(4\pi^2)}
    \right)
  \right] .
\end{equation}
The coupling constants $\alpha$ and $G$ are not independent
parameters for brane world models considered here. 
Substituting Eqs. (\ref{G1:GNJL}) and (\ref{G2:GNJL}) into Eq.(\ref{cl:GNJL}),
we find the critical coupling constant $\alpha_c$.
In Tab. \ref{GNJLTab} we summarize values of the
critical coupling constant for some characteristic cases. Since the
local four-fermi approximation must be valid for heavy KK
excitations, we consider the case $R \sim 1/\Lambda$ for the flat
extra dimension and $kb_0\sim \pi$ for the warped extra dimension 
where only the lightest KK mode has mass below the cut-off scale. 
In the flat extra dimension the lightest KK excitation mode is 
almost decoupled. Thus, the KK modes
have only a small correction to enhance chiral symmetry
breaking. The result is similar to that without KK modes, i.e. QED. On
the other hand, the critical coupling in warped extra dimension is less
than a half value of that in QED in the $y=1/2$ brane, because the coupling constants between the fermion and the KK modes are enhanced by the mode expansion function $\hat{\chi}(y^\ast)$.

\begin{figure}[htbp]
\begin{center}
\centerline{
\epsfig{figure=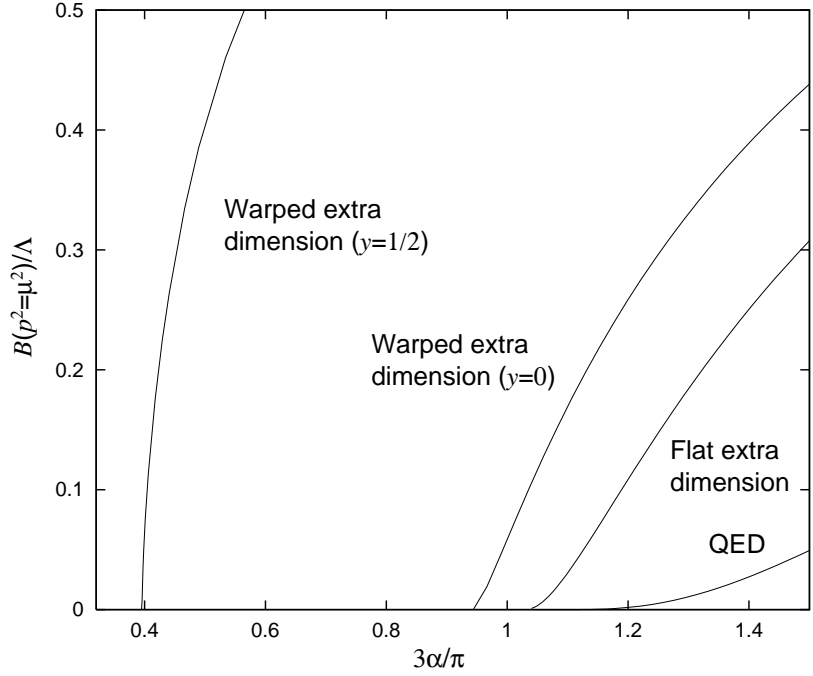,width=0.66\linewidth}
}
\end{center}
\caption{Behavior of the mass function $B(p=\mu)$ as a function
 of $\alpha$ for QED, flat extra dimension ($R\sim 1/\Lambda$) and the
 warped extra dimension ($kb_0\sim \pi$) in the gauged NJL model.}
\label{Ba0}
\end{figure}

In Fig. \ref{Ba0} we show the behavior of the mass function $B(p^2)$ as
a function of the coupling constant $\alpha$ for a fixed $p$. 
The mass function smoothly
disappears as the coupling constant $\alpha$ decreases. In other words,
no mass gap is observed. Thus, the second
order phase transition is realized.
Such a behavior is known as a generalization of the Miranski scaling.
On the $y=0$ brane in the warped extra dimension the effective coupling
constant of the lightest KK mode is almost $\alpha$, and the situation
is similar to the brane in the flat extra dimension. Therefore, 
in the following calculations, we focus on the y = 1/2
 brane for the warped extra dimension and the y = 0 brane is ignored.

\section{Numerical analysis of the Schwinger-Dyson equation}

In the previous section we assumed that the KK excited modes were heavy
enough, and we studied the gauged NJL model as a low-energy effective theory
on the brane. The four-fermi approximation may be valid for heavy KK 
modes. To examine the validity of the approximation in the previous 
section, we introduce the Schwinger-Dyson (SD) equation \cite{SD}.
\begin{eqnarray}
  iS^{-1}(p)&\equiv& A(-p^2)p\!\!\!/ - B(-p^2)
\nonumber \\
  &=& p\!\!\!/ + \sum_{n=0}^{N_{KK}} \int \frac{d^4 q}{i(2\pi)^4}
      e^2\gamma^{\mu}S(q)D^{(n)}_{\mu\nu}(q-p)
      \Gamma^{(n)}_{\nu}(q,p;q-p) ,
\label{eq:SD1}
\end{eqnarray}
where $S(q)$, $D^{(n)}_{\mu\nu}(q-p)$ represents the full propagators for the fermion and the n-th KK excitation modes, and $\Gamma^{(n)}_{\nu}(q,p;q-p)$ is the full vertex function. To solve this equation, we adopt the ladder approximation in which the propagators, $D^{(n)}_{\mu\nu}(q-p)$, and the vertex parts, $\Gamma^{(n)}_{\nu}(q,p;q-p)$, are approximated by tree forms. Within the ladder approximation, the SD equation reduces to the self-consistent equation for the full fermion propagators, $S(p)$. After some calculations the ladder SD equation reads \cite{Ban}
\begin{eqnarray}
A(p^2) &=& 1+ \int_0^{\Lambda^2} dq^2\, \frac{q^2A(q^2)}{A^2(q^2)q^2 +B^2(q^2)} 
            \sum_{n=0}^{N_{KK}} L_\xi (p^2,q^2;M_n,\alpha_n), 
\label{eq:SDA}
\\
B(p^2) &=& \int_0^{\Lambda^2} dq^2\, \frac{q^2 B(q^2)}{A^2(q^2)q^2 +B^2(q^2)} 
         \sum_{n=0}^{N_{KK}} K_\xi (p^2,q^2;M_n,\alpha_n),
\label{eq:SDB}
\end{eqnarray}
where $\xi$ is a gauge-fixing parameter (see appendix) and
\begin{eqnarray}
L_\xi (p^2,q^2;M_n,\alpha_n)
&=& \frac{\alpha_n}{4\pi} 
    \Bigg[ q^2f_{M_n^2}^2 (p^2,q^2) 
         + \frac{2q^2}{M_n^2} \left\{ f_{M_n^2} (p^2,q^2)-f_{\xi M_n^2} (p^2,q^2) \right\} 
           \nonumber \\  && \hspace{1cm} 
         + \frac{(q^2)^2+p^2q^2}{2M_n^2} \left\{ f_{M_n^2}^2 (p^2,q^2)-f_{\xi M_n^2}^2 (p^2,q^2) \right\} 
    \Bigg], \\
K_\xi (p^2,q^2;M_n,\alpha_n)
&=& \frac{\alpha_n}{4\pi} 
    \bigg[ 4f_{M_n^2} (p^2,q^2) 
         + \frac{p^2+q^2}{M_n^2} \left\{ f_{M_n^2} (p^2,q^2)-f_{\xi M_n^2} (p^2,q^2) \right\} 
           \nonumber \\  && \hspace{1cm} 
         - \frac{p^2q^2}{2M_n^2} \left\{ f_{M_n^2}^2 (p^2,q^2)-f_{\xi M_n^2}^2 (p^2,q^2) \right\} 
    \Bigg],
\end{eqnarray}
and
\begin{eqnarray}
f_M (p^2,q^2) 
&=& \frac{2}{p^2+q^2+M+\sqrt{(p^2+q^2+M)^2 -4p^2q^2}} .
\end{eqnarray}
The modified coupling constant $\alpha_n$ is equal to $\alpha$ in the
flat extra dimension. In the warped extra dimension $\alpha_n$ is given by
\begin{equation}
\alpha_n \equiv \alpha \hat{\chi}_n (y^\ast)^2, \ \alpha_0 = \alpha .
\end{equation}
The function $B(p^2)$ is the traceless part of the fermion self-energy. It corresponds to the fermion mass function. If $B(p^2)$ develops a non-vanishing vacuum expectation value, the fermion acquires the dynamical mass, and the chiral symmetry is broken down dynamically. Thus we can regard the mass function, $B(p^2)$, as the order parameter of chiral symmetry breaking. To study if the fermion acquires a dynamical mass or not, we numerically analyze the simultaneous integral equations (\ref{eq:SDA}) and (\ref{eq:SDB}) in terms of bare quantities. 

First, we analyze the flat 
extra dimension with $R\sim 1/\Lambda$ and the $y=1/2$ brane in the warped
extra dimension with $kb_o \sim \pi$. Since only the lightest KK mode
mass is less than the cut-off scale, it is natural to take $N_{KK}=1$.
To solve the Eqs.~(\ref{eq:SDA}) and (\ref{eq:SDB}) numerically, we
employ the iteration method. Starting from suitable trial functions for
$A(p^2)$ and $B(p^2)$, we numerically evaluate the right-hand sides of
Eqs. (\ref{eq:SDA}) and (\ref{eq:SDB}). We next use the resulting
functions, $A(p^2)$ and $B(p^2)$, to calculate the right-hand sides of
Eqs. (\ref{eq:SDA}) and (\ref{eq:SDB}), and we iterate the calculational 
procedure until stable solutions are obtained. At each iteration, the integration is performed by using the Monte Carlo method, and it is cut off at mass scale $\Lambda$.

\begin{figure}[htbp]
\begin{center}
\centerline{
\epsfig{figure=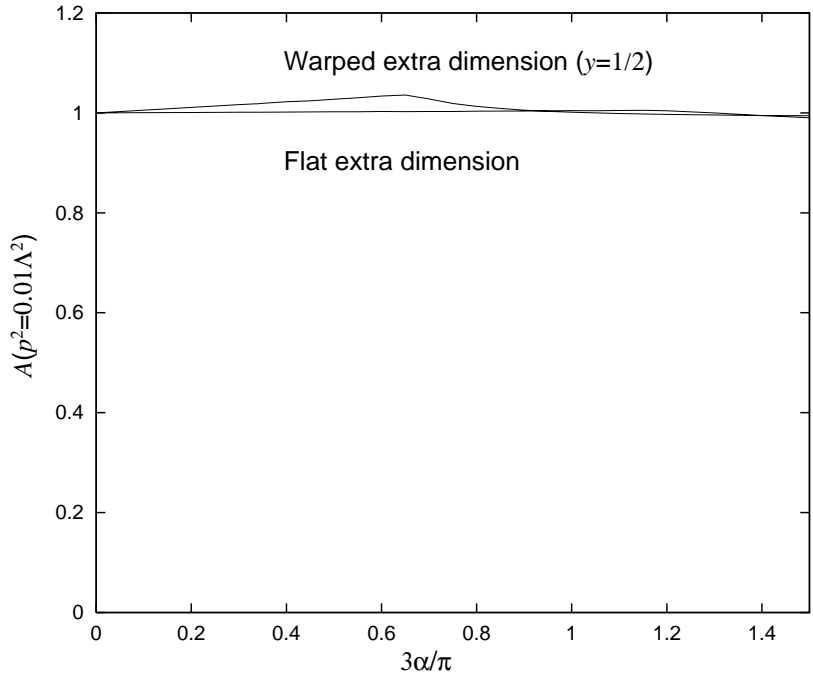,width=0.66\linewidth}
}
\end{center}
\caption{Behavior of the wave function $A(p=0.01\Lambda)$ as a function
 of the coupling constant $\alpha$ for flat extra dimension ($R\sim 1/\Lambda$) with $\xi\sim 0.10$ and the
 warped extra dimension ($kb_0\sim \pi$) with $\xi\sim 0.31$.}
\label{Aa}
\end{figure}

\begin{figure}[htbp]
\begin{center}
\centerline{
\epsfig{figure=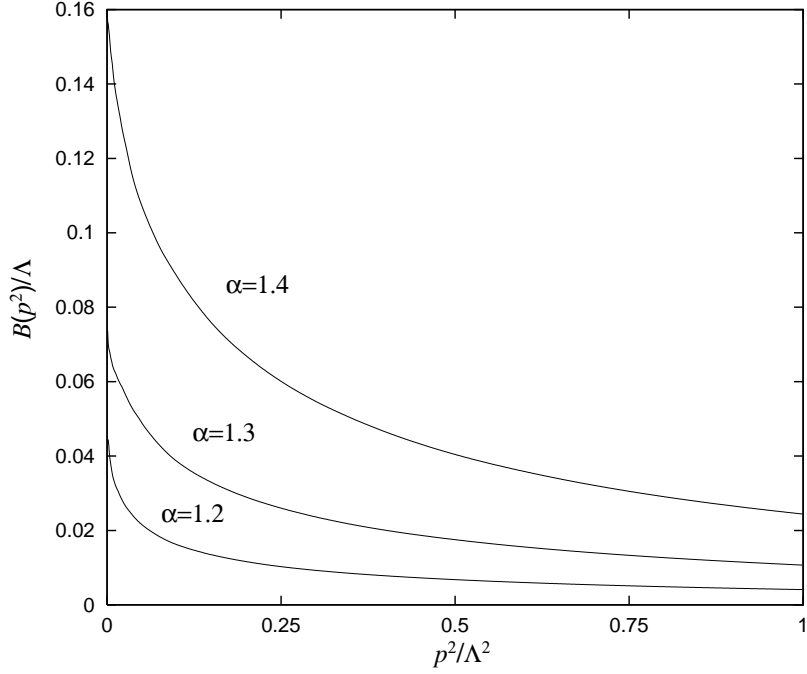,width=0.66\linewidth}
}
\end{center}
\caption{Behavior of the mass function $B(p)$ as a function of the
 momentum $p$ in the flat extra dimension with the coupling constant
 $\alpha$ fixed at $1.2$, $1.3$ and $1.4$.}
\label{Bp:ADD}
\end{figure}

\begin{figure}[htbp]
\begin{center}
\centerline{
\epsfig{figure=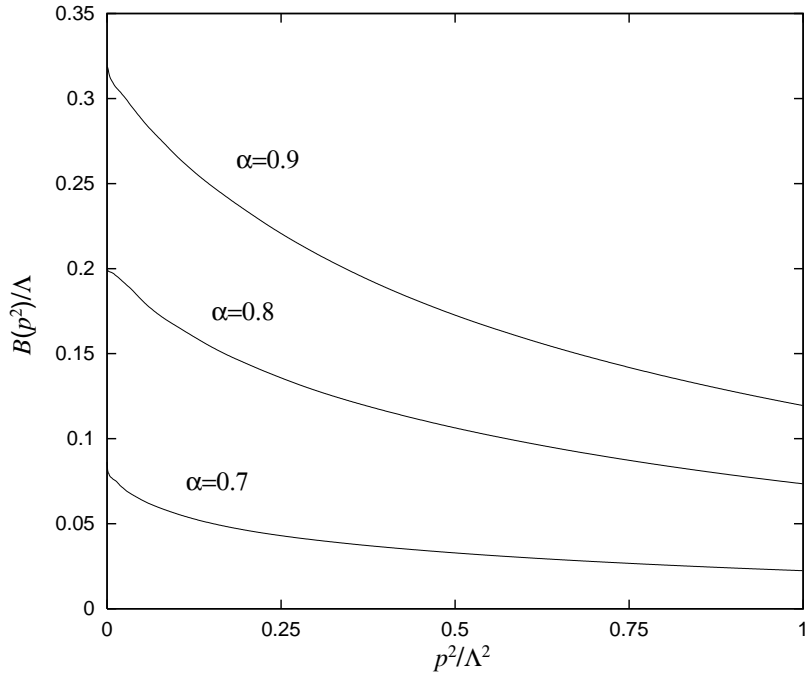,width=0.66\linewidth}
}
\end{center}
\caption{Behavior of the mass function $B(p)$ as a function of the
 momentum $p$ in the $y=1/2$ brane in the warped extra dimension with the coupling constant
 $\alpha$ fixed at $0.7$, $0.8$ and $0.9$.}
\label{Bp:RS}
\end{figure}

If the gauge fields do not propagate in the extra dimension, i.e. no KK
modes exist, the Landau gauge, $\xi=0$, is consistent with the
Ward-Takahashi identity for the ladder SD equation. In the bulk gauge
theories massive KK modes appear in the brane, and the Ward-Takahashi
identity is not guaranteed within the ladder approximation in the Landau
gauge. We try to solve the ladder SD equation by varying the gauge
parameter $\xi$. As is illustrated in Fig. \ref{Aa}, it is possible to
choose the gauge parameter $\xi=\xi_{WT}$ for which the fermion wave function
$A(p^2)$ is almost unity, and the Ward-Takahashi identity nearly holds
on the brane. In the models considered here, $\xi_{WT}$ is given by
\begin{eqnarray}
  \xi_{WT} \sim 0.10,& \mbox{     (for flat extra dimension, }R\sim1/\Lambda\mbox{)}
\label{xi:Trus}
\\
  \xi_{WT} \sim 0.31.& \mbox{     (for warped extra dimension, } kb_0\sim \pi, y=1/2 \mbox{)}
\label{xi:RS}
\end{eqnarray}
Below we choose the gauge parameter $\xi=\xi_{WT}$. In
Figs. \ref{Bp:ADD} and \ref{Bp:RS}, the typical behaviors of
$B(p^2)$ normalized by the cutoff scale $\Lambda$ are presented as a function
of the momentum $p$ for fixed $\alpha$.

\begin{figure}[htbp]
\begin{center}
\centerline{
\epsfig{figure=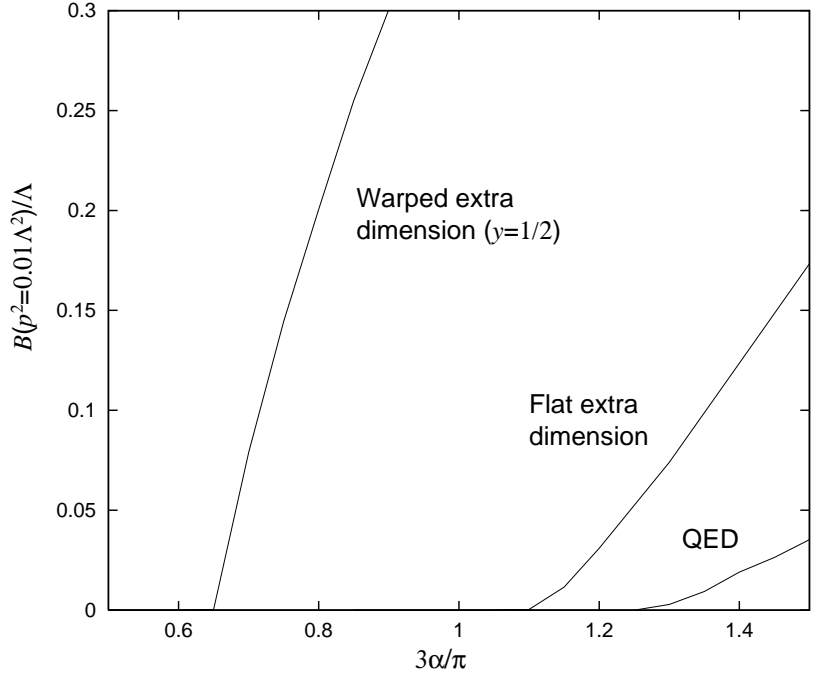,width=0.66\linewidth}
}
\end{center}
\caption{Behavior of the mass function $B(p=0.01\Lambda)$ as a function
 of $\alpha$ for QED, flat extra dimension ($R\sim 1/\Lambda$) and the
 warped extra dimension ($kb_0\sim \pi$) in the ladder SD equation.}
\label{Ba}
\end{figure}

To study the structure of the chiral phase transition, we observe the
behavior of the mass function $B(p^2)$ at some fixed value of $p$. The
$\alpha$ dependence of the mass function $B(p)$ with $p=0.01\Lambda$ is
drawn in Fig. \ref{Ba}. As is seen in Fig. \ref{Ba}, the chiral phase
transition is of the second order (or continuous), since a fermion mass is
generated at a critical value of the coupling constant $\alpha$ without
any discontinuity. For the flat extra dimension, $M^4\otimes S^1/Z_2$, the 
lightest KK mode increases the mass function $B(p)$, but the result has no 
large difference from QED. The result for the warped extra dimension is
far from the other cases. In the $y=1/2$ brane the critical coupling 
$\alpha_c$ is almost half of that in QED. 
It comes from the large effective coupling $\alpha_1 \sim 4\alpha$ for 
the lightest KK mode. Comparing Fig.~\ref{Ba0} with Fig.~\ref{Ba}, we 
observe that the result of the ladder SD equation is qualitatively 
consistent with that obtained by the local four-fermi approximation in the 
previous section.

Next, we study how the number of KK modes, $N_{KK}$, affects the chiral phase 
transition. $N_{KK}$ is defined as the nuamber of KK modes whose mass is less 
than the cut-off scale $\Lambda$. Here we consider the SD equation in the flat 
extra dimension with $R\sim 5/\Lambda, 10/\Lambda, 15/\Lambda, 20/\Lambda$. 
Obviously $N_{KK}$ for each radii is given by $5, 10, 15$ and $20$ respectively. 
In the warped extra dimension KK mode's mass depends on the RS 
compactification scale,
$b_0$, which is defined in Eq.~(\ref{metric}). We also evaluate the SD 
equation in the warped extra dimensions where $N_{KK}$ corresponds to 
$5, 10, 15$ and $20$ each. By solving Eq.~(\ref{eq:KK}) it is found that these
$N_{KK}$ are realized for $kb_0\sim 1.80\pi, 2.22\pi, 2.47\pi$ and $2.65\pi$
respectively. Here we assume that the heaviest KK mode's mass is near the
cut-off scale. 

\begin{figure}[htbp]
\begin{center}
\centerline{
\epsfig{figure=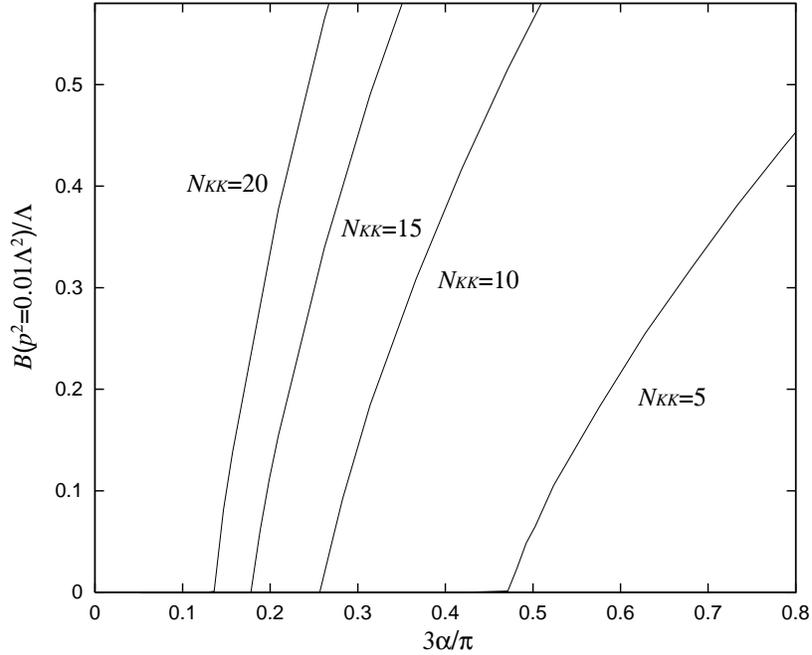,width=0.66\linewidth}
}
\end{center}
\caption{Behavior of the mass function $B(p=0.01\Lambda)$ as a function
 of $\alpha$ for flat extra dimension with
 $R=5/\Lambda (N_{KK}=5)$, $R=10/\Lambda (N_{KK}=10)$, 
 $R=15/\Lambda (N_{KK}=15)$ and $R=20/\Lambda (N_{KK}=20)$
in the ladder SD equation.}
\label{ADDNkk}
\end{figure}

\begin{figure}[htbp]
\begin{center}
\centerline{
\epsfig{figure=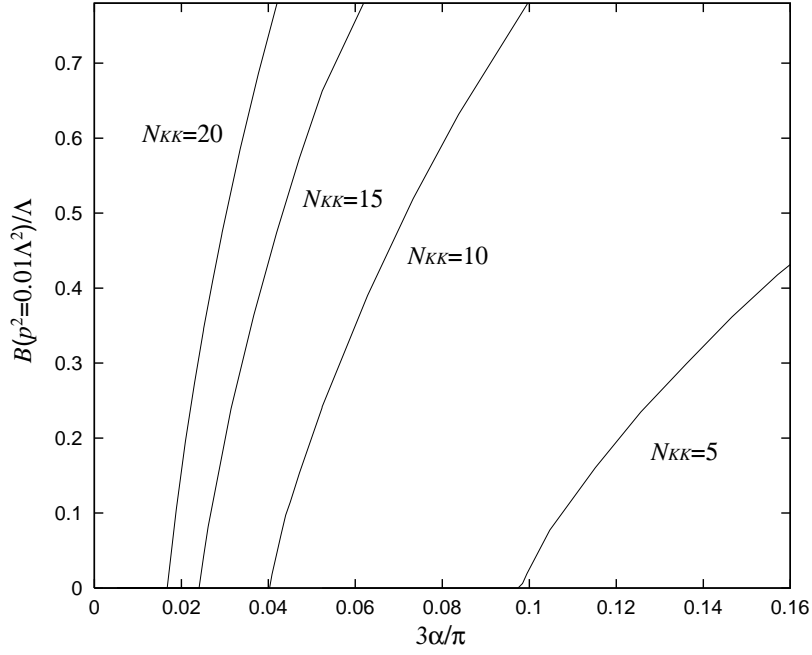,width=0.66\linewidth}
}
\end{center}
\caption{Behavior of the mass function $B(p=0.01\Lambda)$ as a function
 of $\alpha$ for the warped extra dimension $(y=1/2)$ 
 with $kb_0=1.80\pi (N_{KK}=5)$, $kb_0=2.22\pi (N_{KK}=10)$, 
$kb_0=2.47\pi (N_{KK}=15)$ and $kb_0=2.65\pi (N_{KK}=20)$
 in the ladder SD equation.}
\label{RS05Nkk}
\end{figure}

In models with $N_{KK}=5, 10, 15$ and $20$, the gauge parameter for which
the fermion wave function is almost unity is give by $\xi_{WT}\sim 0.27$.
Taking the gauge parameter $\xi=\xi_{WT}$, we solve the Eqs.~(\ref{eq:SDA}) 
and (\ref{eq:SDB}) numerically.
In Figs. \ref{ADDNkk} and \ref{RS05Nkk} we illustrate the behaviors of the 
mass function $B(p^2)$ for the flat extra dimension and for the $y=1/2$ brane 
in the warped extra dimension. In both brane models, chiral smmtery breaking
is enhanced by KK mode summation. For the $y=1/2$ brane in the warped extra 
dimension the critical coupling approaches the ordinary electromagnetic 
coupling constant as $N_{KK}$ increases. 

It is expected that chiral symmetry breaking is extremely enhanced by 
the warp factor for larger $kb_0$. There is a possibility that the chiral 
symmetry is broken down in ordinary electromagnetic coupling constant in 
the warped extra dimension for large $N_{KK}$. 

\section{conclusion}

We have investigated the mechanism of chiral symmetry breaking in
Abelian gauge theories with massless fermion in the brane. We consider
brane worlds in the space-times $M^4\otimes S^1/Z_2$ and $AdS^5$ (RS
model). The induced four-dimensional Lagrangian are constructed from the
five-dimensional theory in the brane world. We adopt the gauge fixing
condition to decouple the extra component of the gauge field. In this
case the contribution of the extra dimension appears as massive KK
excitation modes for ordinary gauge fields in the brane. It is very
important for chiral symmetry breaking that the coupling constant
between KK modes and fermions becomes strong in the $y=1/2$ brane in the
RS model.

To study the influence of the KK modes on the chiral phase transition, we
assume that the KK modes are heavy enough, and consider the gauged NJL
model as a low-energy effective theory. According to the bifurcation
method, chiral symmetry breaking is of the second order. In the
brane included in a flat extra dimension and the $y=0$ brane in a warped
extra dimension, heavy KK modes are almost
decoupled. They have only a small effect on chiral symmetry
breaking. However, it is found that chiral symmetry breaking is
extremely enhanced in the $y=1/2$ brane in the warped extra dimension
due to the large effective coupling. For example, the
critical coupling in the warped extra dimension with $kb_0=\pi$ is
almost half of that in QED in the $y=1/2$ brane. 

In the gauged NJL model, propagators of the KK modes are approximated
by the local four-fermion interactions. To justify this approximation
we consider the SD equation including the effect of the lightest KK mode
propagation. We choose the gauge parameter to hold the Ward-Takahashi
identity, and we numerically solve the ladder SD equation. The result is
qualitatively consistent with that in the gauged NJL model for $N_{KK}=1$.
For the gauged NJL model we pinch the KK modes propagation, and use 
the bifurcation method where the infrared part of the integral is cut 
off at the mass scale $\mu$. Thus, the dynamical fermion
mass and the critical coupling constant is not explicitly equivalent
in both the calculations. 
The effective couplings of the KK modes become stronger and the KK
mode's mass becomes lighter as $kb_0$ increases in the $y=1/2$ brane
in the warped extra dimension. 

The KK modes summation enhanses the chiral symmetry breaking. The critical
value of the coupling constant decreases as $N_{KK}$ increases.
If the chiral symmetry is broken down for ordinary electromagnetic 
coupling constant, all the charged fermions may acquire a dynamical mass. 
But it takes long time to solve the ladder SD equation numerically as 
$N_{KK}$ increases. Furthermore, to estimate the scale of the dynamical 
fermion mass, we must solve the SD equation in terms of renormalized 
quantities. This will be the subject of a forthcoming paper.

The other improvement we should make is to extend our analysis to a
variety of brane world scenarios, and to construct a model of dynamical
symmetry breaking in grand unified theories. It is also interesting to
apply our results to the critical phenomena of electroweak symmetry 
breaking. We will continue our work and we hope to report on these 
problems.

\section*{Acknowledgments}
The authors would like to thank M.~Hashimoto, M.~Tanabashi and T.~Kugo
for fruitful discussions and correspondences.
We also thank Y.~Katsuki, H.~Miguchi, S.~Mukaigawa, T.~Muta and K.~Ohkura for stimulating discussions. 
The numerical analysis was executed on the computer system, parallel pc array for numerical simulations (pan), in Hiroshima University.
T.~I. is grateful for the financial supported provided by the Monbu
Kagakusho fund, Grant-in-Aid for Scientific Research (C) from the
Ministry od Education, Science Sports and Culture, contract number 
11640280.

\appendix
\section{Gauge fixing condition}
In this appendix we present the gauge fixing terms for the bulk Abelian 
gauge field with the brane. We apply the procedure which is known as
$R_\xi$ gauge in the massive gauge theory with the spontaneously broken 
gauge symmetry. First, we consider the common covariant gauge fixing term in 
$(4+d)-$dimensional space-time, with the Lorentz indices 
$M,N=0,1,2,\cdots,3+d$. The term is written as
\begin{eqnarray}
{\cal L}_{\rm GF+FP}
 &=& -i \delta_B \left[ \bar{c} \left( 
        \partial^M A_M + \frac{1}{2} \alpha B \right) \right]
    \nonumber\\ 
 &=& B \partial^M A_M + \frac{1}{2} \alpha B^2 
                      + i \bar{c} \partial^M \partial_M c \nonumber \\
 &=& \frac{1}{2} \alpha \left( B + \frac{1}{2} \partial^M A_M \right)^2 
     - \frac{1}{2 \alpha} \left( \partial^M A_M \right)^2 
     + i \bar{c} \partial^M \partial_M c, 
\label{CGF} 
\end{eqnarray}
where $\delta_B$ represents the BRST transformation, $B$ is the 
Nakanishi-Lautrup field and $\alpha$ is the gauge fixing parameter.
This term is covariant under the full $(4+d)$-dimensional Lorentz 
transformation.

Next we want to consider the case in which there exist 3-branes 
in the direction of the Lorentz index $\mu=0,1,2,3$. 
The 3-brane breaks the Lorentz symmetry. 
In this case the gauge fixing term can be 
written in the form 
\begin{eqnarray}
{\cal L}_{\rm GF+FP}
 &=& -i \delta_B \left[ \bar{c} \left( 
        \partial^\mu A_\mu + \beta \partial^m A_m 
     + \frac{1}{2} \alpha B \right) \right] 
\nonumber \\ 
 &=& B \left( \partial^\mu A_\mu + \beta \partial^m A_m \right)
     + \frac{1}{2} \alpha B^2 
     + i \bar{c} \partial^\mu \partial_\mu c 
     + i \beta \bar{c} \partial^m \partial_m c \nonumber \\
 &=& \frac{1}{2} \alpha \left( B + \frac{1}{\alpha} \partial^\mu A_\mu 
     + \frac{\beta}{\alpha} \partial^m A_m  \right)^2 
     - \frac{1}{2 \alpha} \left( \partial^\mu A_\mu 
     + \beta \partial^m A_m  \right)^2 \nonumber \\ && 
     + i \bar{c} \partial^\mu \partial_\mu c 
     + i \beta \bar{c} \partial^m \partial_m c, 
\label{BGF} 
\end{eqnarray}
where $\beta$ is another gauge fixing parameter, $\mu,\nu=0,1,2,3$ 
and $m,n=4,5,\cdots,3+d$.
The second term in the last line of Eq. (\ref{BGF}) is similar to the 
usual $R_\xi$ gauge which cancels out the mixing with Nambu-Goldstone 
boson. After eliminating the Nakanishi-Lautrup field B, we obtain 
\begin{eqnarray}
{\cal L}_{\rm GF+FP}
 &=& - \frac{1}{2 \alpha} \left( \partial^\mu A_\mu 
     + \beta \partial^m A_m  \right)^2 
     + i \bar{c} \partial^\mu \partial_\mu c 
     + i \beta \bar{c} \partial^m \partial_m c \nonumber \\ 
 &=& - \frac{1}{2 \alpha} \left( \partial^\mu A_\mu \right)^2 
     - \frac{\beta}{\alpha} 
       \left( \partial^\mu A_\mu \right) \left( \partial^m A_m \right)
     - \frac{\beta^2}{2 \alpha} \left( \partial^m A_m \right)^2 
       \nonumber \\ 
&&     + i \bar{c} \partial^\mu \partial_\mu c 
     + i \beta \bar{c} \partial^m \partial_m c. 
\label{GFwB} 
\end{eqnarray}
On the other hand, the Lagrangian for the bulk Abelian gauge field 
contains the following terms:
\begin{eqnarray}
-\frac{1}{4}F^{MN}F_{MN} 
 &=& \frac{1}{2} A_M \left( \eta^{MN} \partial^2 
             -  \partial^M \partial^N \right) A_N \nonumber \\
 &=& \frac{1}{2} A_\mu \left( \eta^{\mu\nu} \partial^2 
             -  \partial^\mu \partial^\nu \right) A_\nu 
     -A_\mu \partial^\mu \partial^m A_m 
\nonumber \\ 
 &&     +\frac{1}{2} A_m \left( \delta^{mn} \partial^2 
             -  \partial^m \partial^n \right) A_n, 
\label{RL} 
\end{eqnarray}
where $\partial^2 = \partial^\mu \partial_\mu + \partial^m \partial_m $.
If we choose $\beta=\alpha$, the second term in Eq. (\ref{GFwB}) and 
that in Eq. (\ref{RL}) are canceled out. The mixing terms between $A_m$ and
$A_{\mu}$ disappear. Thus, the extra components $A_m$ and the Faddeev-Popov 
ghosts $c$ and $\bar{c}$ are decoupled from the brane electrodynamics.

Up to here we have treated the flat space-time case, however we can
easily extend it
to the curved background case, such as the Randall-Sundrum space-time, 
by replacing the derivatives with the general coordinate covariant 
derivatives. In this paper we use the gauge fixing terms discussed here 
with gauge fixing parameter $\xi=1/\alpha$.

\begin{table}[htbp]
\begin{center}
\begin{tabular}{c|c|c|c|c|c}
   \        &   $N_{KK}$ & Radius &   $M_1$    & $\alpha_1$ & $\alpha_{c}$ \\
\hline \hline
QED         & 0 &      \         &       \        & 0  & $\pi/3 \sim 1.05$       \\
Flat extra dim.       & 1 & $R \Lambda =1$  &   $\sim\Lambda$  & $\alpha$  & $1.03$                  \\
Flat extra dim.       & $\infty$ 
            & $R \Lambda =1$  &   $\sim\Lambda$    & $\alpha$ & $1.01$                  \\
Warped extra dim. (y=1/2)         & 1 & $kb_0=\pi$         &
 $\sim0.80k$  & $\sim 4.0\alpha$ & $0.40$                    \\
Warped extra dim. (y=0)         & 1 & $kb_0=\pi$         &   $\sim0.80k$  & $\sim 0.9\alpha$ & $0.94$                    
\end{tabular}
\end{center}
\caption{The critical couplings $\alpha_{c}$ evaluated from 
the result of the four-fermi approximation for $N_{KK}=1$. 
$R$ is the radius of the flat extra dimension. 
The radius of warped extra dimension, $r$, is defined as $b_0 \equiv 2 \pi R$, and $k=\Lambda$.
In the third line, $N_{KK} \to \infty$ in the flat 
extra dimension, meaning that the summation of the KK mode is taken 
to be infinity but $\Lambda$ is fixed to be finite. This corresponds to an 
anisotropy between the cut-off of the bulk and of the brane.} 
\label{GNJLTab}
\end{table}

\end{document}